# Secular variations in the solar corona shape according to observations during a solar activity minimum epoch

A. G. Tlatov

*Kislovodsk Solar Station of the Main Astronomical Observatory, RAS*
tlatov@mail.ru

**Abstract.** Analysis of the solar corona structure during the periods of minimum solar activity from 1867 till 2006 has been carried out. A new flattening index for the large coronal streamers has been proposed. It has been shown that the index has been smoothly changing during the last 140 years. The minimal value of the index occurred during activity cycles 17--19; this was the period when the solar corona most of all corresponded to the dipole configuration of the global magnetic field of the Sun. At the beginning of the 20th and the 21st centuries, the corona structure corresponded to the quadrupole configuration. The reasons for the variations in the solar corona structure and its relation to geomagnetic activity are discussed.

## 1. Introduction

The solar corona structure corresponds to the configuration of solar magnetic fields. Since the magnetic field of the Sun is subjected to cyclic variations, the corona shape also changes cyclically. Processing 12 photographs of the corona during solar eclipses, *Ganskiy* [1897] classified 3 types of solar corona, i.e., maximal, intermediate, and minimal. In 1902, in the report concerning the solar eclipse of 1898, *Naegamvala* [1902] also gave the corona classification depending on the sunspot activity. Description of the corona shape involves the use of characteristic features and the phase of solar activity $\Phi$ which is given by

$$\Phi = \frac{T - T_{\min}}{|T_{\max} - T_{\min}|}.$$

The values of $\Phi$ are positive and negative at the rising and declining branches of the solar cycle. *Vsekhsvyatskiy et al.* [1965] gave a somewhat differing classification of the structure types. They are (1) a maximal type $|\Phi| > 0.85$ in which polar ray structures (PRSs) are not seen, large streamers are observed at all heliolatitudes and are situated radially; (2) an intermediate premaximum or postmaximum type $0.85 > |\Phi| > 0.5$, in which PRSs are observed at least in one hemisphere, large coronal streamers situated almost radially are clearly seen at high latitudes; (3) an intermediate preminimum or postminimum type $0.5 > |\Phi| >$

0.15 in which PRSs are clearly seen in both hemispheres and large coronal streamers strongly deviate toward the solar equator plane; (4) a minimal type 0.15 > |Φ| in which PRSs are clearly seen in both hemispheres and large coronal streamers are parallel to the equator plane; and (5) an ideally minimal type 0.05 > |Φ| in which powerful structures of large coronal streamers are situated along the equator. Changes in the extent of the PRSs, the degree of corona flattening, the average angle between large coronal streamers and other characteristics of the corona depending on the solar cycle phase have been widely studied [*Loucif and Koutchmy,* 1989; *Vsekhsvyatskiy et al.*, 1965].

This paper discusses variations in the corona shape with the phase of the secular activity cycle.

## 2. Processing Method and Results

The initial data in the analysis were drawings of the corona shape taken from the catalogues [*Loucif and Koutchmy,* 1989; *Naegamvala,* 1902; *Vsekhsvyatskiy et al.*, 1965] and also drawings of the eclipses at minima of cycles 23 and 24 taken from *Gulyaev* [1998] and Internet. *Vsekhsvyatskiy et al.* [1965] separated out an ideally minimal corona type. It is supposed that the cycle phase must be |Φ| < 0.05. In fact, the ideally minimal corona type was observed once, in 1954 (Figure 2). Note that this happened at the solar activity minimum before cycle 19, which was the largest during the history of observations. Most probably, such a corona type did not occur during 50 years before and after this event. The corona shape of 1954 was close to a dipole one. This means that large coronal streamers rapidly approach the solar equator plane. At the same time, during other eclipses, such as 1 January 1889 (Φ = -0.18), 21 December 1889 (Φ = 0.03), 17 May 1901 (Φ = -0.07), 10 September 1923 (Φ = 0.04), and 30 May 1965 (Φ = 0.14), the large coronal streamers at distances more than 2 radii do not tend to the equator plane, they propagate along it. In order to analyze the corona shape at the eclipses during a solar activity minimum epoch, a corresponding index should be chosen. We need the index that characterizes the shape of the corona of the minimal type and is applicable to images and drawings of different qualities. The corona of the solar minimum is characterized by pronounced polar ray structures and large coronal streamers. Let us introduce index $\gamma$ that characterizes the angle between high-latitude boundaries of the large coronal streamers at a distance of 2R. The $\gamma$ index is the sum of the angles at the eastern and western limbs: $\gamma = \gamma_W + \gamma_E$. Figure 1 gives the scheme showing how parameter $\gamma$ is determined**.** In fact, the $\gamma$ index is a simpler version of the corona flattening indices [*Ludendorff,* 1928; *Nikolskiy,* 1955].

Figure 2 presents the shapes of eclipses for the epochs close to the solar activity minima of cycles 11--24. Calculated values of parameter $\gamma$ for these eclipses are listed in Table 1. The $\gamma$ parameter varies within 80 – 170 degrees.

Table 1 also gives the solar cycle phase. Figure 3 shows changes in parameter γ for the corona of the solar activity minimum epoch with the phase |Φ| < 0.2. One can see that the lowest magnitudes of the parameter occurred during the period 1935--1955. The remaining magnitudes fit fairly well the enveloping curve with a minimum during cycles 17--19. No information on the solar corona structure during the eclipses at the minima of cycles 18 and 22 has been found in literature. To fill the gaps, the eclipses during the phases of growth or decline of the solar cycle can be used. One can see in Table 1 that the eclipses of 1945 and 1984 are rather far from the solar activity minimum phase. A modified parameter γ*=180 - γ(1-|Φ|) can be introduced for these eclipses. This parameter reduces parameter γ to the phase of minimum. Figure 4 presents variations in parameter γ* during the last 13 activity cycles.

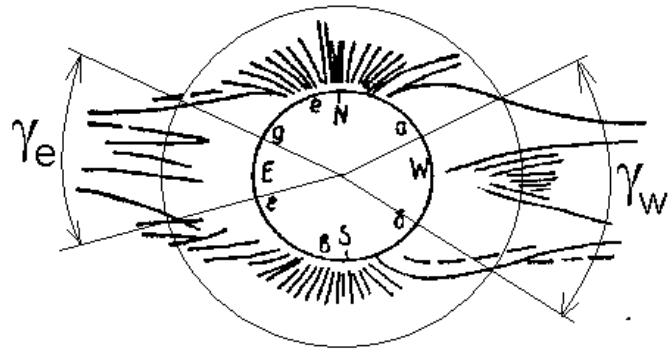

Figure 1. Scheme showing how the angles defining parameter γ = γ$_W$ + γ$_E$ are found for the eclipse of 1923.

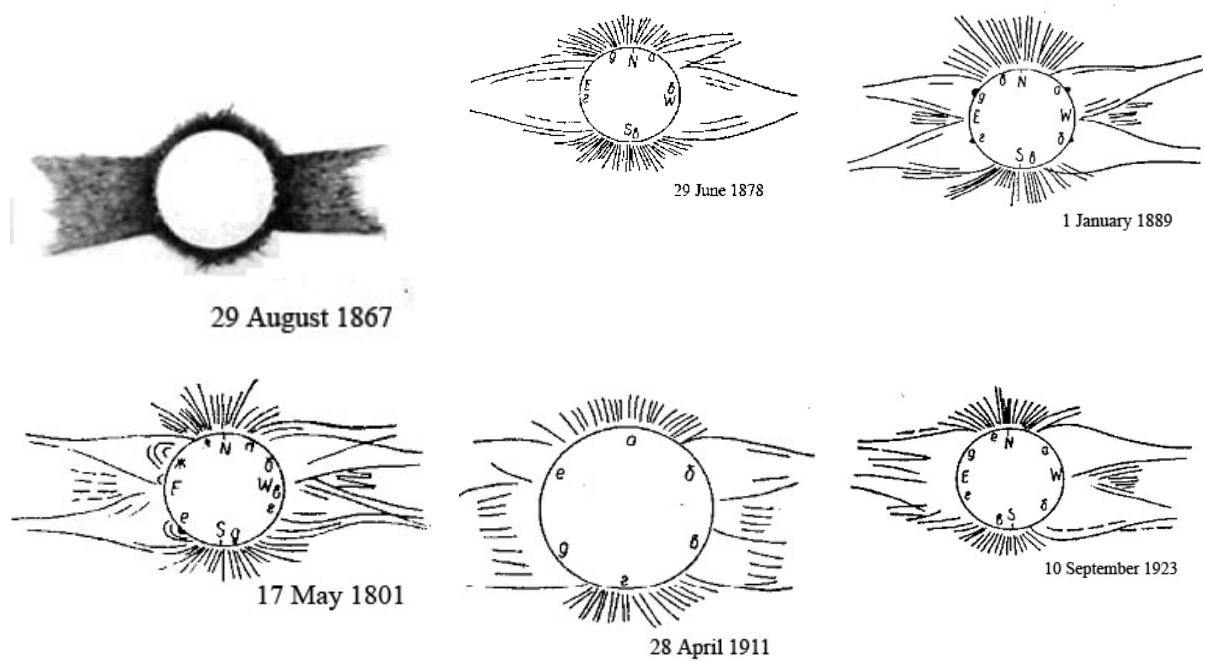

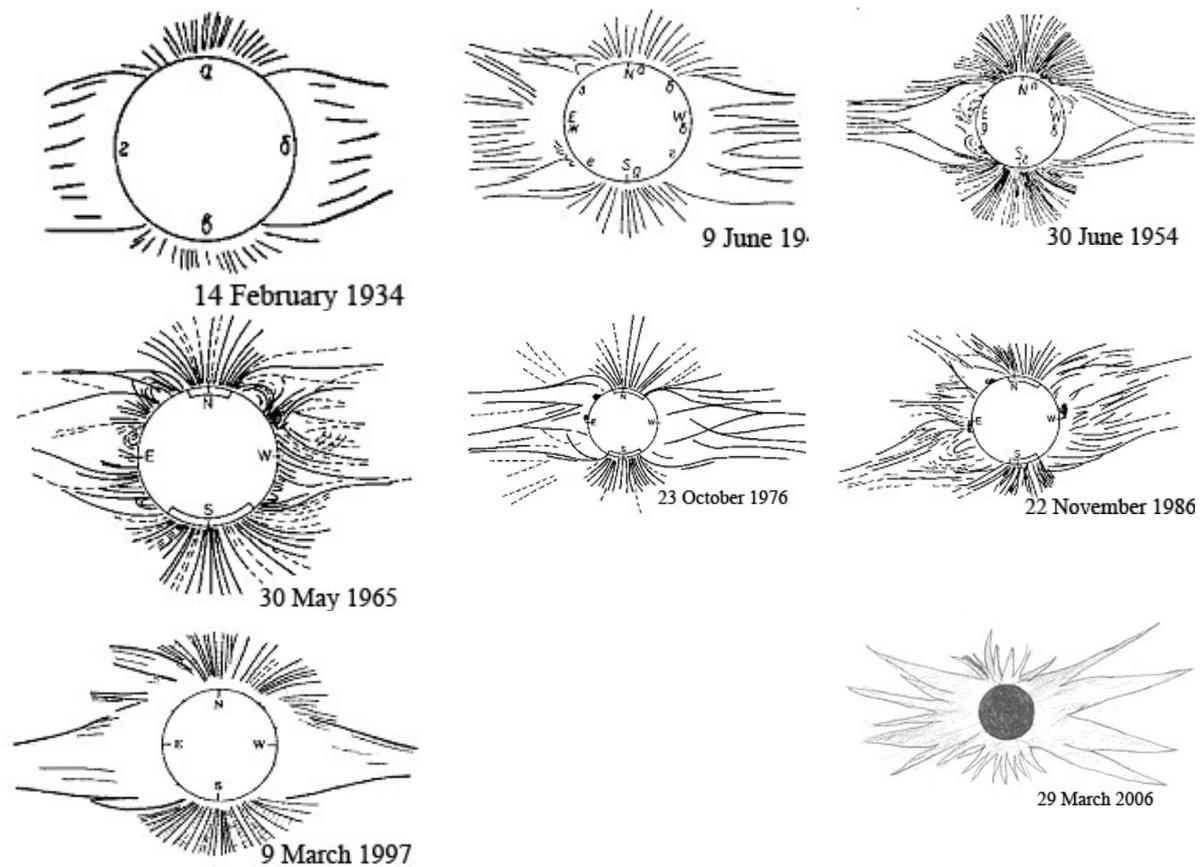

Figure 2. Eclipses close to the minima of cycles 11--24.

The presence of long-term trends in the solar corona structure can be caused by changes in the configuration of the global magnetic field of the Sun. The role of active formation during a solar activity minimum is not significant. It is known that large coronal streamers lie typically above the polarity-inversion lines of the large-scale magnetic field marked by filaments and protuberances [*Vsekhsvyatskiy et al.*, 1965]. For this reason, investigations of the corona shape give valuable information on the structure of the large-scale fields during a long time interval. During the activity minimum, the properties of the global magnetic field of the Sun manifest themselves in the most pronounced way. As a rule, its

Table 1. Parameters γ and γ* = 180 - γ for the Eclipses of Cycles 11 – 24.

| Cycle No | Date | W | γ | Φ* | γ*= 180-γ |
|---|---|---|---|---|---|
| 11 | 29 August 1867 | 140 | 97 | 0.1 | 83 |
| 12 | 29 June 1878 | 75 | 115 | -0.06 | 65 |
| 13 | 21 December 1889 | 88 | 120 | 0.03 | 60 |
| 14 | 17 May 1901 | 63 | 100 | -0.07 | 80 |
| 15 | 28 April 1911 | 103 | 100 | -0.18 | 80 |
| 16 | 10 September 1923 | 77 | 95 | 0.04 | 85 |
| 17 | 14 February 1934 | 114 | 83 | 0.14 | 97 |
| 18 | 09 June 1945 | 151 | 122 | 0.28 | 88* |
| 19 | 30 May 1954 | 190 | 82 | 0 | 98 |
| 20 | 30 May 1965 | 106 | 105 | 0.14 | 90 |
| 21 | 23 October 1976 | 155 | 130 | 0,08 | 48 |
| 22 | 22 November 1984 | 158 | 170 | -0.35 | 69* |
| 23 | 09 March 1997 | 125 | 120 | 0.1 | 60 |
| 24 | 29 March 2006 | -- | 135 | ~0.1 | 40 |

For the eclipses of cycles 18 and 22 parameter γ was reduced to the epoch of minimum as γ* = 180 - γ(1-|Φ|). The cycle phase Φ and Wolf numbers W in the subsequent activity cycle are also given.

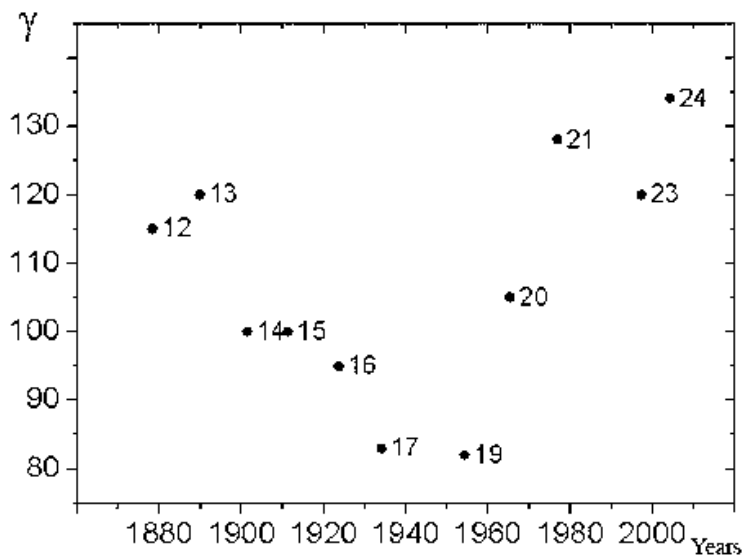

Figure 3. Distribution of parameter *γ* for the structure of the corona of the minimal type with the phase less than |Φ| < 0.2. Numbers of activity cycles are given.

dipole component associated with the growth in the magnetic field strength at the poles is meant. Along with this, one can conclude from the analysis that the assumption that the global solar field configuration is in the form of a dipole structure is probably incorrect. The corona configurations for the eclipses of 1889, 1901 and others correspond rather to the quadrupole form, or octopole form if different polarities at the poles are taken into account.

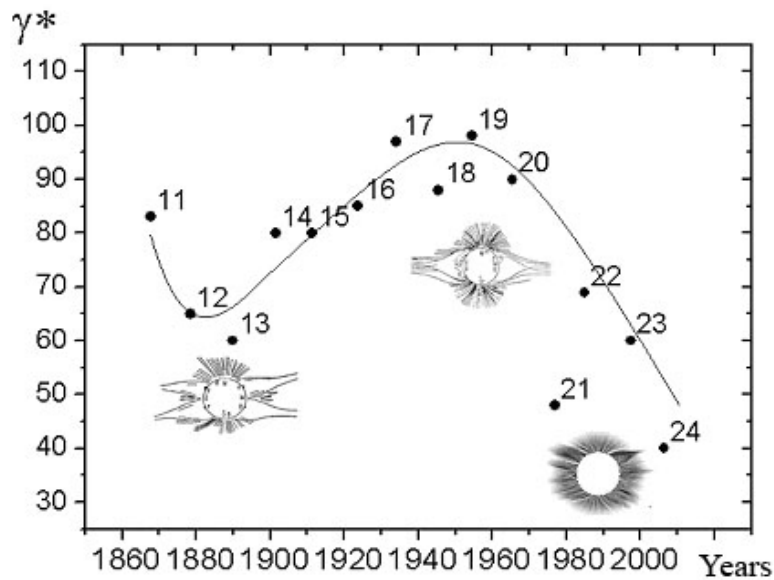

Figure 4. Distribution of modified parameter γ*. The eclipses of cycles 18 and 22 and the enveloping line are also shown.

Thus, long-term variations in parameter γ* should manifest changes in the dipole component during solar activity minima. This hypothesis can be checked using the data on configurations of the large-scale magnetic fields. Figure 5 shows changes in the dipole moment and the envelope drawn through solar activity minima. The data were obtained from the analysis of synoptic H-alpha charts of patterns of polarity inversion lines. The greatest dipole moment corresponded to the minimum of cycle 19 in 1954. On the whole, the enveloping line of the dipole moment corresponds to the changes in the corona shape index γ*.

The growth in the strength of the radial component of the interplanetary magnetic field [*Cliver and Ling,* 2002] determined from the geomagnetic activity index aa is another important problem which has been widely discussed recently. Figure 6 shows variations in the geomagnetic aa index taken from *Cliver and Ling* [2002] The first half of the 20th century was characterized by a growth in this index, i.e., the slowly varying component that was especially pronounced during the activity minimum epochs grew. During the last decades, a decrease in the aa index during the activity minimum epochs was observed. Probably, this is due to rearrangement of the global magnetic field of the Sun accompanied by changes in the solar corona structure during the minimum epochs. Note that variations in the geomagnetic index and the dipole moment of the large-scale magnetic field of the Sun during the activity minimum epoch are almost identical (Figure 7).

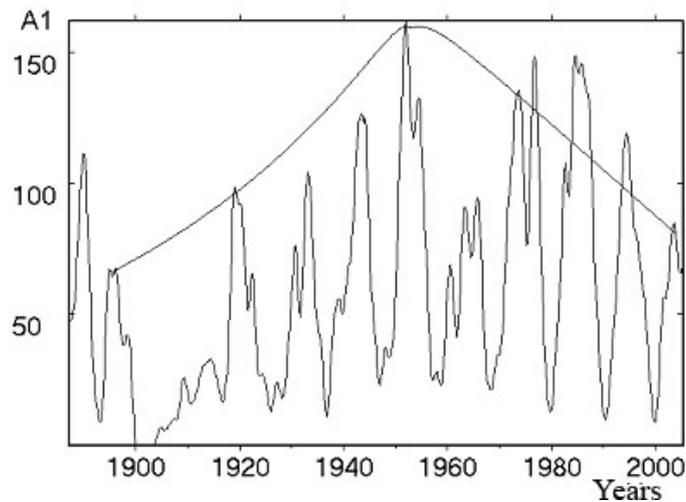
Figure 5. Variations in the dipole moment derived from synoptic H-alpha charts of the Sun. The enveloping line for the solar activity minima is plotted.

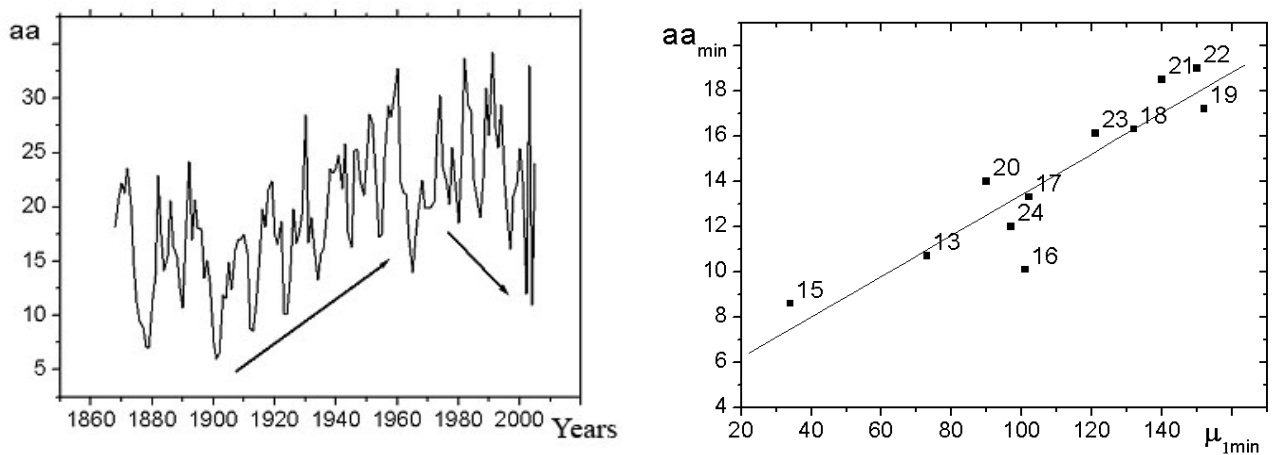
Figure 6. (Left) Annual mean aa indices from 1868. The arrows mark the index growth in the first half of the 20th century and the index decrease during the last decades for the solar activity minimum epochs.
Figure 7. (Right) Relation between aa index and the magnitude of the dipole moment of the large-scale magnetic field µ during the solar activity minimum epoch. Numbers of activity cycles and linear regression are also shown.

Thus, analysis of the corona shape has revealed a long-term modulation of the global magnetic field of the Sun. Possibly, there exists a secular modulation of the global solar magnetic field which is most pronounced during the solar activity minimum epoch. During the secular cycle of the global magnetic field of the Sun the relation between the dipole and octopole components of the magnetic field changes. The largest amplitude of the dipole component occurred during the interval 1944--1955. At the boundary between the 20th and 21st centuries the solar corona shape and, possibly, the global magnetic field correspond to the configuration close to the octopole one. This conclusion does not contradict the calculations of the inclination angles of polar ray structures from the data on the large-scale fields with a maximum in 1940--1950 [*Obridko and Shelting,* 1997]. Note that the maximum of the secular variation in the global magnetic field of the Sun occurred before cycle 19 and preceded the sunspot activity maximum. This allows us to put forward the hypothesis that secular variations in the solar activity are caused by secular modulation of the global magnetic field of the Sun. Another conclusion of this work is the supposition that the slowly changing component of the geomagnetic activity derived from the data on the aa index is due to changes in the dipole component of the large-scale field of the Sun (Figure 7).

**Acknowledgments**. The work was supported by the Russian Foundation for Basic Research (project No. 06-02-16333) and the Program "Nonstationary Processes in Astronomy".